\documentclass[aps,prd,twocolumn,epsf,showpacs,superscriptaddress]{revtex4}
\usepackage{graphicx}% Include figure files
\usepackage{amsmath,amssymb,latexsym}
\usepackage{bm} % bold math

\def\Tr{\text{Tr}} % to use \Tr instead on \matrhm{Tr}

\begin{document}

\title{Monopoles, abelian projection, and gauge invariance}

\author{Claudio Bonati}
\affiliation{Dipartimento di Fisica and INFN, Pisa, Italy}.
\author{Adriano Di Giacomo}
\affiliation{Dipartimento di Fisica and INFN, Pisa, Italy}.
\author{Luca Lepori}
\affiliation{SISSA and INFN Trieste Italy}.
\author{Fabrizio Pucci}
\affiliation{Dipartimento di Fisica and INFN, Firenze, Italy}.

\begin{abstract}
A direct connection is proved between the Non-Abelian Bianchi Identities(NABI), and the abelian 
Bianchi identities for the 't Hooft tensor. As a consequence the existence of a non-zero 
magnetic current is related to the violation of the NABI's and is a gauge-invariant property. 
The construction allows to show that not all abelian projections can be used to expose monopoles 
in lattice configurations: each field configuration with non-zero magnetic charge identifies its 
natural projection, up to gauge transformations which tend to unity at large distances. It is 
shown that the so-called maximal-abelian gauge is a legitimate choice. It is also proved, 
starting from the NABI, that monopole condensation is a physical gauge invariant phenomenon, 
independent of the choice of the abelian projection.
\end{abstract}
\maketitle

\section{Introduction}
Monopoles have become very popular in $QCD$ after the proposal \cite{'tHP, m} that they can
play a role in the confinement of color, by condensing in the vacuum and producing dual 
superconductivity. 

The prototype monopole configuration is a soliton solution of the $SU(2)$ Higgs model with the 
Higgs in the adjoint representation \cite{'tH, Pol}. It exists in the Higgs broken phase
and is characterized by its topology, namely by being a non trivial mapping of the sphere at 
spatial infinity, $S_2$, onto the gauge group, better onto $SU(2)/U(1)$. In mathematical 
language it is said that it has a non trivial $\pi_2$. Similar configurations also exist in the 
unbroken phase, and in a theory with no Higgs field. In this case it is not clear what plays the 
role of the Higgs field and how the unitary gauge is defined. One possibility is that the 
residual $U(1)$ symmetry is identified by the invariance group of any operator 
in the adjoint representation \cite{'tH2}, or, more generally, by any gauge fixing. 
A choice of the residual symmetry is called an abelian projection.

A big activity has developed during the years in the lattice community, to detect and study 
monopoles in numerically generated $QCD$ configurations. After a gauge fixing the abelian part 
of the elementary Wilson loops (the plaquettes) is isolated and monopoles detected by the recipe 
of Ref.\cite{dgt}: any excess over $2\pi$ of the abelian phase is interpreted as existence of a 
Dirac string through the plaquette; a net magnetic charge  exists in an elementary cube when a 
net number of Dirac  flux lines  crosses the plaquettes at its border. In the $U(1)$ gauge 
theory of Ref.\cite{dgt} the magnetic flux is a well defined, gauge invariant quantity 
and so is the number of monopoles.  In the non-abelian case, instead, the abelian part of the 
plaquette is gauge dependent, and therefore the existence or non existence of a monopole in a 
given site is a gauge dependent statement. Indeed most people prefer to work in the so-called 
maximal abelian gauge, in which the abelian part of the phase of the plaquette is maximized, 
and abelian and monopole dominance are direct \cite{sch, suz, pol}. It is rather intriguing that 
a monopole exists in a given site of a configuration in some gauge, and not in some other gauge. 
To have physical meaning a monopole should have a gauge invariant status. Either monopoles can 
be created or destroyed by  gauge transformations, and therefore they are not gauge invariant 
objects, or what is seen on the lattice are mostly artifacts which should disappear at 
sufficiently large $\beta = { 2N \over g^2}$ where the continuum is reached, the lattice spacing 
goes to zero and the average phase of the plaquette with it. In this limit the existence of 
monopoles should be a gauge invariant property. The alternative is that the idea of 
Ref.\cite{'tH2}, that any gauge fixing can provide an effective Higgs field to expose monopoles, 
is not correct.
 
In this paper we address these questions and the problem of giving monopoles a gauge invariant 
definition. The key tool to do that will be the non abelian Bianchi identities.

\section{Non-abelian Bianchi identities}
We define a current $J_{\nu}$ as
\begin{equation}
J_{\mu} =D_{\nu}G^*_{\mu \nu} \label{nabi}
\end{equation}
with $G^*_{\mu \nu} \equiv {1\over 2} \epsilon_{\mu \nu\rho\sigma}G_{\rho \sigma}$ and 
$G_{\rho \sigma}$ the usual field strength tensor. A non zero $J_{\nu}$ is by definition a 
violation of the non-abelian Bianchi identities. Eq.(\ref{nabi}) is gauge covariant and therefore 
a non-zero  $J_{\nu}$ is a gauge-invariant property.
  
In this paper  we will show  that the (abelian) magnetic currents related to monopoles are 
proportional to $ J_{\nu}$, and therefore a violation of the Bianchi identities Eq.(\ref{nabi}) 
is a necessary and sufficient condition for the existence of magnetic currents in a field 
configuration.
  
We will also argue that a natural and general way exists to define the unitary gauge or the 
abelian projection in any field configuration. This choice does not rely on any arbitrary 
choice of a field in the adjoint representation.
      
The current  $J_{\nu}$ is covariantly conserved
\begin{equation}
D_{\nu}J_{\nu} =0 \label{cd}
\end{equation}
Indeed
\begin{eqnarray*}
&& D_{\nu}J_{\nu}={1\over 2} [D_{\mu},D_{\nu}]G^*_{\mu \nu}\propto\\
&& \hspace{1cm}\propto[G_{\mu \nu},G^*_{\mu \nu}] = {1\over 2} \epsilon_{\mu \nu\rho \sigma} 
[G_{\mu \nu},G_{\rho \sigma}]=0
\end{eqnarray*}
The last equality follows from the fact that the fields are taken at the same point, the 
commutator is antisymmetric under the exchange $(\mu \nu)\rightleftarrows (\rho \sigma)$ whilst 
$\epsilon_{\mu \nu\rho \sigma}$ is symmetric.      
      
Not all of the information contained in Eq.(\ref{nabi}), however, is gauge invariant. To identify
its gauge invariant content we can e.g. diagonalise it by a gauge transformation: there will be 
as many independent equations as the number of independent diagonal generators in the algebra. 
This number is by definition the rank $r$ of the group. So for example for $SU(N)$ there will be 
$N-1$ independent equations instead of the $N^2-1$ equalities of Eq.(\ref{nabi}). 
      
To expose them one can project on a complete set of diagonal matrices, which we shall choose 
for convenience to be the matrices $\phi^a_0$ representing the fundamental weights $\mu^a$ 
$(a=1,..,r)$ of the algebra in the representation in which they are diagonal. In formulae:
\begin{equation}
\Tr(\phi^a_0 [D_{\mu} G^*_{\mu \nu}]_{\mathrm{diag}}) = \Tr(\phi^a_0[J_{\nu}]_{\mathrm{diag}}) 
\label{a1}
\end{equation}
In the generic gauge, if we define $\phi^a$ as
\begin{equation}
\phi^a=U(x)\phi^a_0U^{\dagger}(x) \label{orb}
\end{equation}
and
\begin{equation} 
G^*_{\mu\nu} =U(x) [G^*_{\mu\nu}]_{\mathrm{diag}}U(x)^{\dagger} 
\end{equation}
Eq.(\ref{a1}) reads
\begin{equation}
\Tr(\phi^a D_{\mu} G^*_{\mu \nu}) = \Tr(\phi^aJ_{\nu}) \label{a2}
\end{equation}
Eq.(\ref{a2}) is the full gauge-invariant content of Eq.(\ref{nabi}), provided that $\phi^a$ as 
defined by Eq.(\ref{orb}) are diagonal in the gauge in which $J_{\nu}$ is diagonal. The violation 
of the non-abelian Bianchi identities selects a  particular choice of the effective Higgs field, 
i.e. of the abelian projection. Lorentz invariance insures, together with the theorems of 
Ref.\cite{CM}, that the four components of the current $J_{\nu}$ can be made diagonal 
simultaneously in color space. 

However, in the spirit of Ref.\cite{'tH2}, we can assume that any choice of $\phi^a$ of the form 
Eq.(\ref{orb}) is valid and consider the identity
\begin{equation}
\partial_{\mu}\Tr(\phi^a G^*_{\mu\nu}) = \Tr(\phi^a D_{\mu}G^*_{\mu\nu}) + 
\Tr(D_{\mu}\phi^a G^*_{\mu\nu}) \nonumber
\end{equation}
with $\phi^a$ any operator of the form Eq.(\ref{orb}). Making use of Eq.(\ref{cd}) we get
\begin{equation}\label{be}
\partial_{\mu} \Tr(\phi^a G^*_{\mu \nu}) -\Tr(D_{\mu}\phi^a G^*_{\mu \nu}) = \Tr(\phi^aJ_{\nu})
\end{equation} 
We will show in  Section IV that the expression on the left side of Eq.(\ref{be}) is nothing but 
the divergence of the dual of the $a$-th 't Hooft tensor. By definition the 't Hooft tensor is a 
gauge invariant tensor equal to the abelian field strength in the representation in which 
$\phi^a$ is diagonal. In formulae
\begin{equation}
\partial_{\mu} F^{a*}_{\mu \nu} = \Tr(\phi^aJ_{\nu})\label{bbe}
\end{equation}
The magnetic current of the $a$-th monopole species is proportional, for any choice of $\phi^a$, 
Eq.(\ref{orb}), to the violation of the non abelian Bianchi identity Eq.(\ref{nabi}). The choice 
of $\phi^a$ is nothing but the choice of the abelian projection.

Eq.(\ref{bbe}) says then that the existence of a non-zero magnetic current is a gauge invariant 
property, independent of the choice of the abelian projection. This is of course true also in 
the special case  discussed above where $\phi^a$  is diagonal with the current $J_{\nu}$.

\section{The 't Hooft tensor}
We briefly recall some basic facts about the 't Hooft tensor, which will be needed in the 
following; for a complete discussion see \cite{'tH, abprog, DLP}.

The 't Hooft tensor $F_{\mu\nu} $ is a gauge invariant tensor that in the unitary gauge 
coincides with the abelian field strength of the residual $U(1)$ gauge group.
 
In a gauge theory with a compact and semi-simple gauge group $G$, there exist as many monopole 
abelian charges as the rank of the group $r$ and correspondingly $r$ independent 't Hooft 
tensors $F_{\mu\nu}^a$. The index $a$, $(a=1,\ldots,r)$, runs on the simple roots  of the 
Lie algebra of $G$. They have the form \cite{DLP}:
\begin{eqnarray} \label{rt}
F_{\mu \nu}^a = \Tr ( \phi^a G_{\mu \nu} ) - \frac{i}{g}\sum_{I}\frac{1}{\lambda_I^{a} } 
\Tr \left( \phi^a [D_{\mu} \phi^a, D_{\nu} \phi^a ] \right) \nonumber\\
- \frac{i}{g} \sum_{I \neq J}\frac{1}{\lambda_{I}^{a} \lambda_{J}^{a}} 
\Tr \left( \phi^a [[D_{\mu} \phi^a, \phi^a], [D_{\nu} \phi^a,\phi^a ]]\right) - \ldots 
\end{eqnarray}
Here $\phi^a (x)$ is defined by Eq.(\ref{orb}), with $\phi_0^a $ the fundamental weights 
corresponding to  the simple roots $\vec \alpha ^a$ $(a= 1,\ldots,r)$. To define the numbers 
$\lambda_{I}^{a}$ appearing in Eq.(\ref{rt}) we recall the relations which define the 
fundamental weights,
\begin{eqnarray}
[\phi_0^a , H_i] &=& 0  \nonumber  \\
\left[\phi_0^{a }, E_{\vec \alpha}\right] 
&=& ({\vec c}^{\,a}\cdot \vec \alpha)E_{\vec \alpha} \label{alg}
\end{eqnarray}
where $H_i  (i=1,\ldots,r)$ are the commuting elements of the Cartan algebra and 
$E_{\vec \alpha}$ the generators corresponding to the roots ${\vec \alpha}$. The vector 
${\vec c}^{\,a}$ corresponding to the simple root $a$ obeys the following  relations with the 
generic simple root ${\vec \alpha}^b$
\begin{equation}
{\vec c}^{\,a}\cdot{\vec \alpha}^b =\delta_{ab} \label{wgt}
\end{equation}
In order to simplify the notation we define
\begin{equation}
c^a_{\vec \alpha} \equiv ({\vec c}^{\,a} \cdot {\vec \alpha})
\end{equation}
Since any positive root $\vec \alpha$  is a sum of integer  multiples of positive roots, 
the product of $\vec c^{\,a}\cdot\vec \alpha$ is nothing but the number of times that the simple 
root  $\vec \alpha^a$ enters in the sum. The coefficients $\lambda_{I}^{a}$ in Eq.(\ref {rt}) are
the different values that $(\vec c^{\,a}\cdot\vec\alpha)^2$ takes as $\vec \alpha$ runs on the 
ensemble of roots, each value taken only once. There will be as many terms in Eq.(\ref{rt}) as 
many different values of $\lambda^a_{I}$. For  $SU(N)$ group, all the simple roots contribute 
with a factor zero or $1$ to any positive root, so that $\lambda$ assumes the only value 
$\lambda=1$ and the tensor reads \cite{'tH, abprog}:
\begin{equation}\label{tH}
 F_{\mu\nu}^a = \Tr ( \phi^a G_{\mu\nu} ) -
{i \over g} \Tr ( \phi^a [D_{\mu} \phi^a, D_{\nu} \phi^a ])
\end{equation}
It is easy to show that the definition Eq.(\ref{rt}) does in fact define the 't Hooft tensor, 
by a simple extension of the argument developed in Ref.\cite{abprog} for the case of $SU(N)$. 
Call $X_2(A,\phi^a)$ the sum of the terms bilinear in the gauge fields  $A_{\mu} A_{\nu}$ in the 
expression Eq.(\ref{rt}).
\begin{eqnarray}
 &X_2&(A,\phi^a)= ig \Tr( \phi^a(\left[A_{\mu},A_{\nu}\right])  \nonumber \\
+&ig& \sum_{I} {1\over \lambda_{I}^a}\Tr (\phi^a \big[ [A_{\mu},\phi^a ], 
[A_{\nu},\phi^a ] \big] ) \nonumber \\
+&ig& \sum_{I\neq J} {1\over {\lambda_{I}^a  \lambda_{J}^a}}\Tr(\phi^a \big[ [[A_{\mu},\phi^a ],
\phi^a ], [[ A_{\nu},\phi^a ],\phi^a]\big]) \nonumber\\
&+& .... \label{x2}
\end{eqnarray}
Going to the gauge where $\phi^a $ is diagonal, $\phi^a =\phi_0^a $ and expanding $A_{\mu}$ on 
the Cartan basis of the algebra $H_i, E_{\vec \alpha}$ we get
\begin{equation}\label{alg2}
A_{\mu} = \sum_{i}a^i_{\mu} H_i + \sum_{\vec \alpha}a^{\vec \alpha}_{\mu} E_{\vec \alpha}
\end{equation}
Because of the cyclic property of the trace only the non diagonal part of $A_{\mu}$ contributes 
in Eq.(\ref{x2}). By use of Eq.(\ref{alg}) the commutators can be computed. The result is
\begin{eqnarray}
X_2(A,\phi^a) = ig\sum_{\vec \alpha}a^{\vec \alpha}_{\mu}a^{-\vec \alpha}_{\nu}\Tr(\phi^a_0
[E_{\vec \alpha},E_{-\vec \alpha}])\times\hspace{1cm} \nonumber \\
\times\left[ 1 - \left(\sum_{I} {1\over {\lambda_{I}^a}}\right)(c^a_{\vec \alpha})^2 
+\left(\sum_{I \neq J} {1 \over {\lambda_{I}^a \lambda_{J}^a}}\right) (c^a_{\vec \alpha})^4  
-\ldots  \right] \label{x22}
\end{eqnarray}
For each root $\vec \alpha$,  $(c^a_{\vec \alpha})^2$ will be equal to some $\lambda_{I}^a$. It 
is trivial to check that the sum in the second line of Eq.(\ref{x22}) is equal to zero , for any 
number of $\lambda$'s.

It follows that
\begin{equation}
 X_2(A,\phi^a)=0 \label{x23}
 \end{equation}
for arbitrary fields $A_{\mu}$.  

For  convenience we now give a name, $\Delta_{\mu \nu}^a$, to the sum of the terms that follow 
the first one in the expression Eq.(\ref{rt}).
\begin{eqnarray} 
\Delta_{\mu \nu}^a = -\frac{i}{g} \sum_{I} \frac{1}{\lambda_I^{a}}\Tr (\phi^a [D_{\mu} \phi^a, 
D_{\nu} \phi^a ]) \hspace{1cm}\nonumber\\  
- \frac{i}{g} \sum_{I \neq J}\frac{1}{\lambda_{I}^{a}\lambda_{J}^{a}}\Tr\left(\phi^a \big[[D_{\mu} 
\phi^a, \phi^a],[D_{\nu} \phi^a,\phi^a ]\big]\right)
- \ldots \label{del}
\end{eqnarray}

In the new notation $F_{\mu \nu}^a$ can be written in a compact form as 
\begin{equation}
F_{\mu \nu}^a = \Tr( \phi^a G_{\mu \nu} ) + \Delta_{\mu \nu}^a \label{abb}
\end{equation}

Noticing that, by use of Eq.(\ref{orb}), $\partial_{\mu} \phi^a =ig[\Omega_{\mu},\phi^a]$  with 
\begin{equation}
\Omega_{\mu} = {1\over {ig}} \partial_{\mu} U(x)U(x)^{\dagger}\nonumber
\end{equation}
we can write
\begin{equation}
D_{\mu} \phi^a = ig [\Omega_{\mu} + A_{\mu}, \phi^a]\label{comm}
\end{equation}
By  use of Eq.(\ref{comm}) and of the definition  of $X_2(A,\phi^a)$ in Eq.(\ref{x2}) we get
\begin{eqnarray}
\Delta_{\mu \nu}^a &=& X_2(A+\Omega,\phi^a) -ig\Tr( \phi^a [A_{\mu}+\Omega_{\mu},A_{\nu}+ 
\Omega_{\nu}] ) \nonumber 
\\ &=& -ig \Tr( \phi^a( [A_{\mu}+\Omega_{\mu},A_{\nu} +\Omega_{\nu}] ) \label{del2}
\end{eqnarray}
 Because of Eq.(\ref{x23}), we have then for $F^a_{\mu \nu}$, 
\begin{eqnarray}
F^a_{\mu \nu}= \Tr(\partial_{\mu} (\phi^a A_{\nu}) -\partial_{\nu}(\phi^a A_{\mu})) \nonumber \\
-ig\Tr(\phi^a [\Omega_{\mu}, \Omega_{\nu}])\hspace{1cm}
 \label{aaa}
\end{eqnarray}
In the unitary gauge $\phi^a$ is equal to the fundamental weight, $\Omega_{\mu}=0$ and 
$F^a_{\mu \nu}$ is nothing but the abelian field strength of the residual $U(1)$ gauge symmetry. 
We conclude that our definition Eq.(\ref{rt}) \cite{DLP} gives the correct 't Hooft tensor.

For gauge groups with one single value for $\lambda_{I}^a$, namely $1$, $\Omega_{\mu}$ in 
Eq.(\ref{aaa}) can be replaced by $[\Omega_{\mu},\phi^a]$ since the commutator of $\phi^a$ with 
the elements of the algebra corresponding to positive roots reproduces the elements, the 
commutator with the elements corresponding to negative roots reproduces them up to a minus sign. 
The trace selects products of elements corresponding to a root and to its negative. As a 
consequence
\begin{eqnarray}
F^a_{\mu \nu} = \Tr(\partial_{\mu} (\phi^a A_{\nu}) -\partial_{\nu}(\phi^a A_{\mu})) \nonumber \\
-{i\over g}\Tr(\phi^a\left[\partial_{\mu}\phi^a, \partial_{\nu}\phi^a\right])\hspace{1cm}
\label{cc}
\end{eqnarray}
Eq.(\ref{cc}) was first derived in Ref.\cite{arafune} for $SU(2)$. See also Ref.\cite{abprog}.
 
For a detailed derivation of Eq.(\ref{rt}) we refer to Ref.\cite{DLP}.

\section{Proof of Eq.(\ref{bbe})}
In order to derive Eq.(\ref{bbe}),  we sum to both sides of the equality Eq.(\ref{be})  the 
divergence of the dual of the tensor $\Delta_{\mu \nu}^a$ defined in Eq.(\ref{del}). Making use 
of Eq.(\ref{abb}) the result reads
\begin{equation}
\partial_{\mu} F^{a\,*}_{\mu \nu} = \Tr( \phi^a J_{\nu} ) + R_{\nu}^a
\end{equation} 
with
\begin{equation}
R_{\nu}^a = \Tr ( D_{\mu} \phi^a G^*_{\mu \nu} ) + \frac{1}{2} \epsilon_{\mu\nu\rho\sigma}
\partial_{\mu} \Delta_{\rho \sigma}^a \hspace{0.3cm}.
\label{defR}
\end{equation}
 We now prove the following\vspace{0.3cm}

\textbf{Theorem:} If $\phi^a$ is any element of the orbit of a fundamental weight $\phi_0^a$ 
Eq.(\ref{orb}), then $R_{\nu}^a = 0$.
\vspace{0.3cm}

If this is true, then $j_\nu^a = \Tr( \phi^a J_{\nu}) $ is equal to the violation of the abelian 
Bianchi identites $\partial_{\mu}  F^{a\,*}_{\mu \nu} = j_\nu^a $, $a=(1,..,r)$, of the 't Hooft 
tensors.

We start by splitting for convenience the second terms of the r.h.s. of Eq.(\ref{defR}) in two 
terms 
\begin{equation} 
\frac{1}{2} \epsilon_{\mu \nu \rho \sigma} \partial_{\mu} \Delta_{\rho \sigma}^a = B_{\nu}^a + 
C_{\nu}^a \label{BC} 
\end{equation}
$B_{\nu}^a $ collects the terms with a double covariant derivative of the field $\phi^a$, 
$C_{\nu}^a$ those with only first order covariant derivatives:
\begin{eqnarray}
&B_{\nu}^a&= -\sum_I \frac{i}{g \lambda_I^a}\epsilon_{\mu \nu \rho \sigma} \Tr (\phi^a 
[ D_{\mu} D_{\rho} \phi^a, D_{\sigma} \phi^a ] ) \nonumber  \\ 
&- &\sum_{I\neq J} \frac{i}{g \lambda_I^a \lambda_J^a}\epsilon_{\mu \nu \rho \sigma} \Tr \left(
\phi^a \big[ [\phi^a, D_{\mu} D_{\rho} \phi^a], [\phi^a, D_{\sigma} \phi^a ] \big] \right) 
\nonumber \\
&-&\ldots \label{defB}
\end{eqnarray}
and
\begin{eqnarray}
&{}&C_{\nu}^a = - \sum_I \frac{i}{2 g \lambda_I^a}\epsilon_{\mu \nu \rho \sigma} 
\Tr( D_{\mu}\phi^a [ D_{\rho} \phi^a, D_{\sigma} \phi^a ])  \nonumber \\ 
&{}&-\sum_{I\neq J} \frac{i}{2 g\lambda_I^a\lambda_J^a}\epsilon_{\mu \nu \rho \sigma}
\Tr\left( D_{\mu} \phi^a \big[ [\phi^a,  D_{\rho} \phi^a],
[\phi^a, D_{\sigma} \phi^a ] \big] \right)  \nonumber \\
&{}&-\sum_{I\neq J} \frac{i}{ g\lambda_I^a \lambda_J^a}\epsilon_{\mu\nu\rho\sigma} 
\Tr\left( \phi^a \big[ [ D_{\mu}\phi^a,  D_{\rho}\phi^a],
[\phi^a, D_{\sigma} \phi^a ] \big] \right) \nonumber\\ 
&{}& -\ldots \label{defC}
\end{eqnarray}

Let's now consider the first term of the r.h.s. of (\ref{defR}), 
$\Tr( D_{\mu} \phi^a G^*_{\mu \nu})$. Since it gets non zero contribution  only from the 
components of the field strength  $G_{\rho \sigma}$ that do not commute with $ \phi^a $ we can 
introduce the projector $P^a$ on these components in an analogous way as was done in 
Ref.\cite{DLP}, and replace $G_{\rho \sigma}$ by $P^aG_{\rho \sigma}$,
\begin{eqnarray*}
P^a G_{\rho \sigma} = \sum_I \frac{1}{\lambda_I^a} [\phi_a,[\phi_a, G_{\rho \sigma}]] \hspace{1cm}\\
-\sum_{I\neq J} \frac{1}{\lambda_I^a \lambda_J^a}  [\phi^a, [\phi^a, [\phi^a,[\phi^a, 
G_{\rho \sigma}]] + \ldots
\end{eqnarray*}
Since
\begin{eqnarray*}
\epsilon_{\mu \nu \rho \sigma} [\phi^a, G_{\rho \sigma}] = \frac{i}{g}\epsilon_{\mu \nu \rho \sigma}  
\big[[D_{\rho}, D_{\sigma}],\phi^a\big] \nonumber \\
= \frac{2 i}{g}\epsilon_{\mu \nu \rho \sigma}  D_{\rho} D_{\sigma} \phi^a \hspace{1.5cm}
\end{eqnarray*}
it follows that
\begin{eqnarray*}
\frac{1}{2} \epsilon_{\mu \nu \rho \sigma} P^a G_{\rho \sigma} =  \frac{i}{g}\epsilon_{\mu \nu \rho 
\sigma} \Bigg\{ \sum_{I }\frac{1}{\lambda_I^a}\ [\phi_a,D_{\rho} D_{\sigma} \phi_a] \\ 
- \sum_{I\neq J} \frac{1}{\lambda_I^a \lambda_J^a}  [\phi^a, [\phi^a, [\phi^a, D_{\rho} D_{\sigma} 
\phi_a]  + ....\Bigg\}
\end{eqnarray*}
and therefore
\begin{eqnarray*}
&& \Tr( D_{\mu}\phi^a P^a G^*_{\mu\nu} ) = \\
&&  +\frac{i}{g} \sum_I \frac{1}{\lambda_I}\epsilon_{\mu \nu \rho \sigma} \Tr (\phi^a [ D_{\mu}
D_{\rho}\phi^a, D_{\sigma}\phi^a ] \\ 
&& +\frac{i}{g} \sum_{I\neq J} \frac{1}{\lambda_I^a\lambda_J^a}\epsilon_{\mu \nu \rho \sigma} 
\Tr\left( \phi^a \big[ [\phi^a, D_{\mu} D_{\rho} \phi^a],[\phi^a, D_{\sigma} \phi^a ] \big] 
\right) \\
&& + \ldots 
\end{eqnarray*}
By comparison with Eq.(\ref{defB})
\begin{equation}
\Tr(D_{\mu}\phi^a P^a G^*_{\rho \sigma}) = -B^a_{\nu}
\end{equation}
Our theorem is then equivalent to the statement
\begin{equation}
C^a_{\nu} =0 \label{Cze}
\end{equation}

The proof of Eq.(\ref{Cze})  that we were able to give is by direct computation. The computation 
is relatively simple for groups with one single value of the coefficients  $\lambda_{I}^a$, like 
$SU(N)$, or with two or three values, like $G_2$, and we report it below. For more complicated 
cases with four to six values of $\lambda_{I}^a$, which occur with the special groups 
$F_4, E_7, E_8$  the computation is more involved, and is sketched in the appendix. The result 
is in all cases Eq.(\ref{Cze}). All the possible patterns are enumerated in  the Table at the 
end Ref.\cite{DLP}. Maybe a more synthetic proof can be given, but we were not able to find it. 
The physically interesting cases involve one value of $\lambda_{I}$ [$SU(N)$], or at most three 
[$G_2$]. The other cases were included, as was done in Ref.\cite{DLP}, only to show that the 
mechanism is very general, relies on geometry, and is independent on the choice of the gauge 
group.
 
For the case of one value of $\lambda_{I}$,  $C_{\nu}^a$ Eq.(\ref{defC}) contains one term 
\begin{equation*}
C_{\nu}^a= -\frac{i}{2g} \epsilon_{\mu \nu \rho \sigma} \text{Tr}\left( D_{\mu} \phi^a [ D_{\rho} 
\phi^a, D_{\sigma} \phi^a ] \right)
\end{equation*}
Using again the expansion  Eq.(\ref{alg2}), in the unitary gauge we obtain 
\begin{eqnarray}
C_{\nu}^a=-\frac{g^2}{2}\sum_{ijk}\epsilon_{\mu \nu \rho \sigma}\, a_{\mu}^{\vec\alpha_i} 
a_{\rho}^{\vec\alpha_j} a_{\sigma}^{\vec\alpha_k} c^a_{\vec \alpha_i} c^a_{\vec \alpha_j} 
c^a_{\vec \alpha_k}
\times\nonumber\\ 
\times \Tr\left( E_{\vec{\alpha}_i} [ E_{\vec{\alpha}_j} , E_{\vec{\alpha}_k}  ] \right)\hspace{1.5cm}
\label{polSU(N)}
\end{eqnarray}
where the sum runs onto all the roots. Since the trace is gauge invariant, this expression is 
nonzero only if there's at least one triple of roots satisfying the relation
\begin{equation} 
\vec{\alpha}_i +\vec{\alpha}_j + \vec{\alpha}_k = 0 \label{inv}
\end{equation}
Moreover each root must  contain the root $\vec \alpha^a$ corresponding to the maximal weight 
$\phi_0^a$ with coefficients plus or minus one, otherwise the coefficients 
$(\vec{c} \cdot \vec{\alpha})$ are equal to zero. The simple roots are linear independent and 
complete.  The combination $\vec \alpha_i +\vec  \alpha_j + \vec \alpha_k $ can never  be zero, 
since it contains the root $\vec \alpha_a$ one or three times.  As a consequence $C^a_{\nu}=0$. 

If there are two different values of $\lambda$ namely $\lambda = 1,4$ we have
\begin{eqnarray}
C_{\nu}^a= -\frac{g^2}{2}\sum_{ijk}\epsilon_{\mu \nu \rho \sigma} a_{\mu}^{\vec\alpha_i} 
a_{\rho}^{\vec\alpha_j} a_{\sigma}^{\vec\alpha_k} c^a_{\vec \alpha_i} c^a_{\vec \alpha_j} 
c^a_{\vec \alpha_k}
\nonumber\times\hspace{1cm}\\
\times\Tr( E_{\vec{\alpha}_i} [ E_{\vec{\alpha}_j} , E_{\vec{\alpha}_k}  ])\times {\Pi_2}\label{ee} 
\hspace{1.0cm}
\end{eqnarray}
\begin{equation}
{\Pi_2} = \sum_I \frac{1}{\lambda_I^a} + \sum_{I\neq J} \frac{1}{\lambda_I^a \lambda_J^a}
\left[c^a_{\vec \alpha_j}c^a_{\vec \alpha_k} - 2 (c^a_{\vec \alpha_i})^2\right] 
 \label{g2-1}
\end{equation}\\
The only possibility to satisfy Eq.(\ref{inv}) is when the simple root ${\vec \alpha}^a$ 
appears once and with the same sign in each of two roots, say $\vec \alpha_j$, $\vec \alpha_k$ 
and twice with opposite sign in the third one $\vec\alpha_i$. The first line of Eq.(\ref{ee}) is 
invariant under permutations of $i,j,k$, so we can sum the second line on the permutations.  
Moreover it follows from Eq.(\ref{inv}) that
$c^a_{\vec \alpha_j}  c^a_{\vec \alpha_k}= {1\over 2}(\lambda_{i} -\lambda_{j}-\lambda_{k})$. 
The average of each $\lambda_{i}$ on the permutations is $\langle\lambda\rangle =2$. The quantity 
$\Pi_2$ in  Eq.(\ref{g2-1})  is finally equal to 
$\Pi_2 = ({5\over 4} - {1\over 4}\cdot{5\over 2}\langle\lambda\rangle) $, which vanishes for 
$\langle\lambda\rangle =2$. This proves the theorem for two values of $\lambda$.
 
If there are three different values for $\lambda$, $1, 4, 9$, Eq.(\ref{defC}) reads
\begin{eqnarray}
 C_{\nu}^a= -\frac{g^2}{2}\sum_{ijk}\epsilon_{\mu\nu\rho\sigma} a_{\mu}^{\vec\alpha_i} 
a_{\rho}^{\vec\alpha_j} a_{\sigma}^{\vec\alpha_k} c^a_{\vec \alpha_i} c^a_{\vec \alpha_j} 
c^a_{\vec \alpha_k} \times \nonumber\\ 
\times \Tr( E_{\vec{\alpha}_i} [ E_{\vec{\alpha}_j} , E_{\vec{\alpha}_k}  ])\times {\Pi_3}\label{eee} 
\hspace{1.5cm}
\end{eqnarray}
\begin{eqnarray}
{\Pi_3} =\sum_I \frac{1}{\lambda_I^a} + \sum_{I\neq J} \frac{1}{\lambda_I^a \lambda_J^a} 
\Big[c^a_{\vec \alpha_j} c^a_{\vec \alpha_k} - 2  (c^a_{\vec \alpha_i})^2 \Big]+ \nonumber\\
+\sum_{I\neq J\neq K} \frac{1}{\lambda_I^a \lambda_J^a \lambda_K^a} \Big[ (c^a_{\vec \alpha_j})^2 
(c^a_{\vec \alpha_i})^2  + 2 (c^a_{\vec \alpha_i})^4 -\nonumber\\
 -2 c^a_{\vec \alpha_k}(c^a_{\vec \alpha_i})^3\Big ] \hspace{1.5cm}
\label{g2-2}
 \end{eqnarray}
Similar to what was done above   $\Pi_3$ can be expressed in terms of the averages over 
permutations $\langle\lambda^a\rangle$, $\langle(\lambda^a)^2\rangle$, 
$\langle\lambda^a_I\lambda^a_J\rangle$ and of the known coefficients 
$\sum_I \frac{1}{\lambda^a_I}={49\over 36}$, 
$\sum_{I\neq J} \frac{1}{\lambda^a_I \lambda^a_J} ={14\over 36}$,
$\sum_{I\neq J\neq K} \frac{1}{\lambda_I^a \lambda_J^a \lambda_K^a} ={1\over 36}$. The result  is
\begin{equation}
\Pi_3 ={49\over36}-{14\over 36}\left({{5}\over{2}}\right)\langle\lambda\rangle +
{1\over 36} (3\langle\lambda^2\rangle+\langle\lambda_i\lambda_j\rangle)
\label{num}
\end{equation}
The possibilities to satisfy Eq.(\ref{inv})  are
\begin{enumerate}
\item $(c^a_i =1, c^a_j=1,c^a_k =-2)$ and permutations 
\item $(c^a_i =1, c^a_j=2,c^a_k =-3)$ and permutations.
\end{enumerate}
and the corresponding values for the averages are
\begin{enumerate}
\item $\langle\lambda\rangle =2$, $\langle\lambda^2\rangle=6$, 
$\langle\lambda_i\lambda_j\rangle =3$,
\item $\langle\lambda\rangle ={14\over 3}$, $\langle\lambda^2\rangle={98\over 3}$, 
$\langle\lambda_i\lambda_j\rangle ={49\over 3}$.
\end{enumerate}
Both choices, when inserted in Eq.(\ref{num}), give $\Pi_3=0$, i.e. $C^a_{\nu}=0$.

The way to extend the proof  to more complicated cases in which $\lambda$ can assume four, five 
or six different values \cite{DLP} is trivial: write $C_{\nu}^a$ in the analogous form as 
Eq.'s (\ref{ee}), (\ref{eee}); identify the triples of roots containing $\vec\alpha^a$ at least 
once, and show by direct computation, as done above, that the sum over permutations of the 
factor $\Pi$ multiplying their contribution to $C_{\nu}^a$ is zero. This is schematically 
presented in the Appendix.

\section{About the abelian projection}
We now check the statements of the previous section on the exact monopole solution of Ref.'s 
\cite{'tH, Pol}. For the notation and for the detailed calculations we refer to the original 
papers \cite{'tH, Pol}, and to the book Ref.\cite{shnir}. In this case the gauge group is 
$SU(2)$, there is one fundamental weight, ${\mu} = {{\sigma_3}\over 2}$ in the representation in 
which it is diagonal. The violation of the non-abelian Bianchi identities of Eq.(1) can 
explicitly be computed, e.g in the unitary gauge, giving, for a monopole sitting at $\vec r=0$,
\begin{equation}
D_iB_i= {{2\pi}\over g} \sigma_3\delta^3(\vec r)\label{nabitHp}
\end{equation}
Here $B_i \equiv {1\over 2} \epsilon _{ijk} G_{jk} $.
 
The current $J_{\nu}$ of Eq.(\ref{nabi}) is diagonal in the unitary gauge, where, by definition, 
the $vev$ of the Higgs field is diagonal, proportional to $\sigma_3$, and therefore the right 
choice for the field $\phi$ of Eq.(\ref{orb}) is  the direction of the Higgs field.
  
For SU(2) a generic operator in the adjoint representation has the form Eq.(\ref{orb}) (up to a 
normalization): a different choice of the abelian projection, i.e. of $U(\vec r)$, as suggested 
in Ref.\cite{'tH2}, gives rise to a different 't Hooft tensor, in particular to a magnetic charge 
renormalized by a factor $\cos(\beta)$, where the angle $\beta$ is the second Euler angle in the 
parametrization $U(\vec 0) =\exp(i{\alpha\over 2}\sigma_{3}) \exp(i{\beta \over 2}\sigma_2) 
\exp( i{\gamma \over 2} \sigma_3)$. The resulting 't Hooft tensor obeys Eq.(\ref{bbe}), with a
magnetic charge renormalized by a factor $\cos(\beta)$ which does not fulfill in general the 
Dirac quantization condition.

Not all of the abelian projections are equivalent, but there exists a natural choice of the 
gauge fixing, namely the unitary gauge.

Notice that the unitary gauge in this model is  nothing but the maximal-abelian gauge: indeed it 
can be directly verified that the gauge field $A_{\mu}$ in the unitary gauge obeys the equation 
\cite{'tH2}
\begin{equation}
\partial_{\mu} A^{\pm}_{\mu} \pm ig \left[A^3_{\mu}, A^{\pm}_{\mu}\right] =0
\label{mag}
\end{equation}
which defines the maximal-abelian gauge. We shall further elaborate on this point below.

Notice that the gauge field of the solution is  (in the hedgehog gauge) 
\begin{eqnarray}     
A^a_0 &=& 0   \nonumber   \\
A^a_{i} & = &   -\epsilon_{aij}{{r^j}\over {gr^2}}[1-K(gvr)]
\end{eqnarray}
Here $g$ is the gauge coupling constant, $v$ the $ vev$ of the Higgs field, and $K(x )$ a 
function whose shape depends on the parameters of the Higgs part of the lagrangian. Up to 
possible logarithmic corrections
\begin{eqnarray}
[1-K(x) ]_{x\to 0} &\propto  &{x}^2        \nonumber \\
K(x)_{x \to \infty}&\propto &\exp(-x)
\end{eqnarray}
The function $K(gvr)$ cancels exactly in the Bianchi identity Eq.(\ref{nabitHp}).
   
If we choose the $\phi$ field diagonal in the unitary gauge $ \phi={\sigma_3\over 2}$, we get 
for the 't Hooft tensor
\begin{eqnarray}
e_i &\equiv &F_{0i} =0 \nonumber \\
b_i & \equiv & {1\over 2} \epsilon _{ijk} F_{jk} = {{r^i}\over {2gr^3}}
\end{eqnarray}
and for the corresponding abelian Bianchi identities
\begin{equation}
\vec \nabla \cdot \vec b ={{2\pi}\over g} \delta^3(\vec r)
\end{equation}
in agreement with the general argument of Section II. The magnetic current has zero space 
components, as expected for a static configuration, and a time independent time component 
Eq.(\ref{nabitHp}) and is therefore trivially conserved. If we look more closely to the 
structure of the 't Hooft tensor, we can separately compute the two terms which enter in its 
definition
\begin{equation}
F_{\mu \nu} = \Tr(\phi G_{\mu \nu}) - {i\over g}\Tr(\phi [D_{\mu}\phi, D_{\nu}\phi]) \label{tt}
\end{equation}
The first term has the form
\begin{equation}\label{th1}
\vec b^{(1)} = {{\vec r}\over {2gr^3}}[1 - K^2] \approx_{r\to \infty} {{\vec r}\over {2gr^3}}
\end{equation}
while the second term has the form
\begin{equation}\label{th2}
\vec b^{(2)}= K^2 {{\vec r}\over {2gr^3}}
\end{equation}
        
At large distances only the first term survives, up to exponentially small corrections. The non 
abelian magnetic field at large distances is in fact abelian and has the same orientation in 
color space as the current $J_{\mu}$ of Eq.(1), which coincides, in this model, with the 
orientation of the vacuum expectation value of the Higgs field. Indeed, in the unitary gauge, 
the non abelian magnetic field $B_i$ at distances much larger than the characteristic length 
${1\over {gv}}$ has the form
\begin{equation}
\vec B= {{\vec r}\over {2gr^3}} \sigma_3 \label{tPB}
\end{equation}
    
These properties are much more general. Consider first a generic static configuration. For such 
configurations it can be shown \cite{coleman} that the magnetic monopole term in the multipole 
expansion of the field at large distances is abelian and, if not equal to zero (zero magnetic 
charge), it selects  a direction in color space. We shall identify this direction  with the 
direction of the field $\phi_0$ as defined in Eq.s (\ref{a1}),(\ref{orb}), or with the direction 
of the Higgs field in the model of Ref.s \cite{'tH, Pol}. This identifies the unitary gauge 
up to gauge transformations $U(\vec r)$ which  tend to the identity as $r\to \infty$.  Such 
transformations do not modify the asymptotic monopole part of the 't Hooft tensor, i.e. the 
magnetic charge of the configuration. The unitary gauge is defined modulo such transformations.   
In this gauge  the magnetic field at large distances will have the form Eq.(\ref{tPB}) with a 
factor ${m \over 2} $ in front, $m$ being the total magnetic charge of the configuration, which 
in Eq.(\ref{tPB}) is equal to 2. The first  term of Eq.(\ref{tt}) gives   
\begin{equation}
\vec b \approx \vec b^{(1)}   \approx_{r\to \infty} {m\over 2}{{\vec r}\over {2gr^3}}\label{asy}
\end{equation}
   
The second term will give a negligible contribution in any case, since, in the unitary gauge, 
$D_{\mu} \phi =ig\left[A^{\pm}_{\mu} ,\sigma_3\right]$, (the diagonal part $A^3_{\mu}$ does not 
contribute) and $A^{(\pm)}_{\mu} $ are non-leading in the multipole expansion. The flux at 
infinity of the abelian magnetic field  corresponds to a magnetic charge of $m$ Dirac units. 
Since the asymptotic field obeys the condition Eq.(\ref{mag}), the maximal abelian 
gauge certainly belongs to the class of unitary gauges defined above.
   
The abelian Bianchi identity Eq.(\ref{bbe}) reads, in the unitary gauge just defined, 
\begin{equation}
\vec \nabla \cdot \vec b \equiv j_0  = \Tr(J_0{ \sigma_3\over 2}) \label{abmax}
\end{equation}
The total magnetic charge 
$Q \equiv \int d^3r j_0(\vec r) $,
is by Gauss theorem applied to Eq.s (\ref{asy}),(\ref{abmax})
\begin{equation}
Q= \frac{m\pi}{g}
\end{equation}
We can now, in the unitary gauge, add and subtract to $\vec b$ a point-like solution with the 
same magnetic charge, Eq.'s (\ref{th1}),(\ref{th2})
\begin{equation}
\vec b = \frac{m}{2}\frac{\vec r}{2gr^3} +\delta \vec {b} \label{asyn}
\end{equation}
with 
\begin{equation}
\delta \vec b = \vec b -  \frac{m}{2}\frac{\vec r}{2gr^3}
\end{equation}
For the magnetic charge density Eq.(\ref{abmax}) we have
\begin{equation}\label{accc}
j_0 = \frac{m\pi}{g} \delta^3(\vec r) + \delta j_0
\end{equation}
where, by construction,
\begin{equation}
\int d^3r  \delta j_0 =0
\end{equation}
Since the 't Hooft tensor is linear in the gauge field, Eq.(\ref{x23}), the procedure amounts to 
add and subtract from  the gauge field that of a classical solution $A^0_{\mu}$ corresponding to 
a point-like  monopole of charge $m$. 
\begin{equation} \label{semcl}
A_{\mu} = A^0_{\mu} + \delta A_{\mu}
\end{equation}
The term $ A^0_{\mu}$ contains a monopole term in the multipole expansion, the term 
$\delta A_{\mu}$ only contains higher multipoles. The non abelian magnetic field $\vec B$ has a 
similar decomposition
\begin{equation}
\vec B = \vec B^0 + \delta\vec B
\end{equation}
and also here the first term will be leading at large $r$. Direct computation of the 't Hooft 
tensor reproduces Eq.(\ref{asyn}). Finally, the non-abelian Bianchi identity Eq.(\ref{bbe}) 
will read
\begin{equation}
D_iB_i =\sigma_3\frac{m\pi}{g} \delta^3(\vec r) + \delta J_0
\end{equation}
After projection on the fundamental weight the first term gives the first term on the right 
side of Eq.(\ref{accc}), while the second one gives $\delta j_0$. The argument is somehow 
self-consistent. 
   
Strictly speaking the theorem of Ref.\cite{coleman} on the asymptotic behavior of the field 
configurations only applies to solutions of the equations of motion. However a  configuration 
which contributes to the Feynman integral in the importance sampling of a lattice simulation 
can be viewed as a solution which carries the topology with quantum fluctuations added to it. 
The classical configuration describes the large $r$ part of the configuration, whilst 
fluctuations have zero average magnetic charge and vanish at large distances. Configurations are 
classified by their $\pi_2$ winding number.
   
This realizes the very idea of duality:  none of the symmetries of $QCD$ is broken, neither in 
the confined and in the deconfined phase, and its equations of motion are obeyed in each sector 
independent of the winding number. The winding number introduces extra degrees of freedom and an 
extra (dual) symmetry, the magnetic $U(1)$ which can explain the phase structure by the 
mechanism of Higgs breaking (magnetic charge condensation). The natural excitations in $(3+1)$ 
dimensions are monopoles.
   
The above argument can be extended to non static configurations. Up to topologically irrelevant 
terms the superposition of the configuration to the time-reversed configuration is static and 
allows to define the magnetic charge. Lorentz invariance allows to define the spatial components 
of the magnetic current.
   
As shown in Ref.\cite{DLP}, the argument can easily be generalized to each monopole species in 
the case of a generic gauge group.
          
As a final remark we notice that the above discussion is relevant  to the detection of monopoles 
in field configurations, e.g. in the lattice configurations. As recalled in the introduction the 
recipe of Ref.\cite{dgt} applied to abelian part of the plaquette in QCD gives different number 
of monopoles, depending on the choice of the abelian projection. Our arguments show that the 
maximal-abelian gauge is the correct unitary gauge, up to gauge transformations which tend to 
unity as $r\to \infty$. With this choice magnetic charges obey the Dirac quantization condition; 
with other choices the magnetic charge is smaller and therefore can prove undetectable by the 
recipe of Ref.\cite{dgt} which is based on Dirac condition.
    
A check of this statement could be that monopoles defined by any abelian projection should also 
show up as monopoles in the maximal abelian projection, at least at sufficiently small coupling 
where possible discretization artifacts should be absent.

\section{Monopole condensation and confinement.}
As shown in Sect.II there are infinitely many abelian conserved currents 
\begin{equation}
j^a_{\nu}\left(x,U\right) = \Tr\left(U(x)\phi^a_0U^{\dagger}(x) J_{\nu}(x)\right)
\end{equation}
one for each choice of $\phi^a$ in Eq.(\ref{orb}), namely one for each conceivable abelian 
projection. $J_{\nu}$ is the violation of the non-abelian Bianchi identities Eq.(\ref{nabi}) and 
is abelian-projection independent.
    
As discussed above one choice is the maximal abelian projection, where $\phi^a$ and $J_{\nu}$ 
commute and the current is given by Eq.(\ref{a1}): let us denote it just by  $j^a_{\nu}$ for the 
sake of notational simplicity. For any other choice of $U(x)$ we have
\begin{equation}
j^a_{\nu}\left(x,U\right) = \Tr\left(U(x)\phi^a_0 U^{\dagger}(x)[J_{\nu}]_{\mathrm{diag}}\right)
\label{aux}
\end{equation}
Since the Lie algebra is isomorphic to the adjoint representation, in the notation of 
Eq.(\ref{alg2}) 
\begin{equation}
U(x)\phi^a_0 U^{\dagger}(x) = \sum_{\vec \alpha} C_{\vec \alpha}^a\left(x,U\right) E_{\vec \alpha} 
+\sum_b C^{ab}\left(x,U\right)\phi^b_0
\end{equation}
Inserting this equality  in Eq.(\ref{aux}) only the diagonal part will contribute to the trace, 
and
\begin{equation}
j^a_{\nu}\left(x,U\right) = \sum_b C^{ab}\left(x,U\right) j^b_{\nu}(x) \label{pcc}
\end{equation}
If $O(y)$ is a local operator carrying a magnetic charge $m$, the commutator of the magnetic 
charge operator  $Q^a\equiv \int d^3x j^a_0(\vec x,t)$ with it will be ($Q^a$ is constant in time)
\begin{equation}
\left[ Q^a, O( y)\right] = m O(y)
\end{equation}
By virtue of Eq.(\ref{pcc}), the commutator of $O(y)$ with the magnetic charge in the abelian 
projection identified by $U(x)$, $Q^a(U) \equiv \int d^3x j^a_{\nu}\left(x,U\right)$ will be
\begin{equation}
\left[Q^a(U),O(y)\right] = C^{aa}\left(x,U\right)mO(y)\label{fund}
\end{equation}
The coefficient $C^{aa}\left(x,U\right)$ will generically be non-vanishing. We are neglecting 
the mixing to other charges  Eq.(\ref{pcc}) which is trivially correct for $SU(2)$ gauge group, 
and is also true if we choose $O(y)$ to be neutral with respect to the charge $Q^b$, 
($b\neq a$). Therefore if  $\langle O\rangle \neq 0$, i.e. if monopoles condense in the 
maximal-abelian gauge, they will condense in any other abelian projection. Somehow the effective 
Higgs is abelian projection independent.
   
This explains the apparent paradox why all quarks are confined, including those which are 
neutral in the maximal abelian gauge. This problem has been discussed since years 
\cite{gw, dmo} and, more recently, in a series of detailed papers Ref.\cite{sco, scol} and 
references therein.
   
All that is a consequence of the non-abelian Bianchi identities Eq.(\ref{nabi}) and of our 
theorem Eq.(\ref{bbe}).

\section{Conclusions}
The basic result of this paper is the connection between magnetic charge and violation of the 
non abelian Bianchi identities. The abelian Bianchi identities of the abelian projected fields 
are obtained from the non abelian ones by projection on the orbits of the fundamental weights 
Eq.(\ref{orb}), Eq.(\ref{bbe}).
    
This analysis allows a critical discussion of the choice of the abelian projection which defines 
the magnetic $U(1)$'s. Any magnetically charged configuration naturally identifies a class of 
abelian projections to expose monopoles, which we show to be the maximal abelian gauge modulo 
gauge transformations which are trivial on the sphere $S_2$ at infinity. In these projections 
monopoles have the correct magnetic charge obeying the Dirac quantization condition.
    
This is relevant if one wants to expose monopoles, e.g. in lattice configurations. For this 
purpose the idea of Ref.\cite{'tH2} to use the direction of any operator in the adjoint 
representation as effective Higgs is inadequate, since the effective magnetic charge is 
smaller than the Dirac charge and can be undetectable by the recipe of Ref.\cite{dgt} which is 
based on Dirac magnetic charge quantization. This can possibly explain why different numbers of 
monopoles are observed, in the same configuration, in different abelian projections. The 
existence of a monopole in a site of a field configuration is in any case a gauge invariant, 
abelian-projection invariant statement, modulo lattice artifacts. 

This aspect deserves a more detailed investigation, specifically on two respects: 
\begin{enumerate}
\item Is the border of an elementary cube of the lattice far enough compared to the 
characteristic length of a monopole to legitimate the use of the asymptotic behavior 
Eq.(\ref{asy})?
\item What is the influence of lattice artifacts affecting the definition of monopole of 
Ref.\cite{dgt} at strong coupling on the detection of monopoles?
\end{enumerate}
   
Lattice artifacts have clearly been observed in Ref.\cite{ddmo}. In the approach of 
Ref.\cite{sco,scol} possibly a way out is provided by the technical recipe of averaging on 
gauge transformed configurations.
    
If instead one wants to create a monopole (Ref.\cite{dp, ddp, ddpp, dlmp}) on a configuration 
with periodic boundary conditions, which has zero magnetic charge, one can choose any  abelian 
projection for the new monopole, or do not even chose it \cite {ccos, 3}. 

As for dual superconductivity, Eq.(\ref{fund}), which is a consequence of Eq.(\ref{bbe}), allows 
us to state that the Higgs breaking of magnetic $U(1)$ is a projection independent property. 
This is why the Abrikosov dual flux -tubes between quarks are projection independent features 
\cite{scol}.    
        
Most of the discussion above refers to the deconfined phase, where the magnetic symmetry is 
realized a' la Wigner, and magnetic charge is well defined as a $U(1)$ generator. We assume, as 
usual, that everything and, in particular  the definition of lattice monopoles, can be extended 
to the confined phase where magnetic charge is not defined due to monopole condensation.
    
\section{Acknowledgments}
One of us (A.~Di~Giacomo) wants to thank P.~A.~Marchetti for enlightening discussions. He also 
thanks M.~Shifman for systematically asking about gauge invariance of monopoles during the years 
at all seminars and meetings on the subject. Discussions with G.~Paffuti are  aknow-ledged. 
F.~Pucci and L.~Lepori thank D.~Seminara and G.~Thompson for useful discussions.

\appendix
\section{Proof of the theorem for exceptional groups}

In this appendix we show how to extend the proof of eq.(8) to the exceptional groups $F_4$, 
$E_7$ and $E_8$, where the $\lambda$ coefficients can take four to six different values 
\cite{DLP}. The $C^a_{\nu}$ tensors read

\begin{eqnarray}
C_{\nu}^a= -\frac{g^2}{2}\sum_{ijk}\epsilon_{\mu \nu \rho \sigma} a_{\mu}^{\vec\alpha_i} a_{\rho}^{\vec\alpha_j} a_{\sigma}^{\vec\alpha_k}
c^a_{\vec \alpha_i} c^a_{\vec \alpha_j} c^a_{\vec \alpha_k}
\nonumber\times\hspace{1cm}\\
\times\Tr( E_{\vec{\alpha}_i} [ E_{\vec{\alpha}_j} , E_{\vec{\alpha}_k}  ])\times {\Pi_{i = 4, 5, 6}}\hspace{1.0cm}
\end{eqnarray}

Here

\small \begin{displaymath}\Pi_4 = \frac{205}{144} - \frac{91}{192}\frac{5}{2}\langle\lambda\rangle+\frac{5}{96}\left( 3 \langle\lambda^2\rangle + \langle\lambda_j \lambda_i\rangle \right)+\end{displaymath}\begin{equation}\, \, \,  +\frac{1}{576}\left( -3 \langle\lambda^3\rangle - 3\langle\lambda_i^2 \lambda_j\rangle + \frac{1}{2} \langle\lambda_i \lambda_j \lambda_k\rangle\right)\end{equation}

\small \begin{displaymath}\Pi_5 = \frac{5269}{3600} - \frac{1529}{2880}\frac{5}{2}\langle\lambda\rangle+\frac{341}{4800}( 3 \langle\lambda^2\rangle + \langle\lambda_j \lambda_i\rangle )+\end{displaymath}\begin{displaymath} +\frac{11}{2880}\left( -3 \langle\lambda^3\rangle - 3\langle\lambda_i^2 \lambda_j\rangle + \frac{1}{2}\langle\lambda_i \lambda_j \lambda_k\rangle\right)\end{displaymath}
\begin{equation} +\frac{1}{14400}(  3 \langle\lambda^4\rangle +2 \langle\lambda_i^2 \lambda_j^2\rangle + 3 \langle\lambda_i^3 \lambda_j\rangle - \langle\lambda_i \lambda_j^2 \lambda_k\rangle)\end{equation}

\begin{displaymath}\Pi_6 = \frac{5369}{3600} - \frac{37037}{64800}\frac{5}{2}\langle\lambda\rangle+\frac{44473}{518400}( 3 \langle\lambda^2\rangle + \langle\lambda_j \lambda_i\rangle )\end{displaymath}\begin{displaymath} +\frac{1001}{172800}\left( -3 \langle\lambda^3\rangle - 3\langle\lambda_i^2 \lambda_j\rangle + \frac{1}{2}\langle\lambda_i \lambda_j \lambda_k\rangle\right)\end{displaymath}
\begin{displaymath} +\frac{91}{518400}(  3 \langle\lambda^4\rangle +2 \langle\lambda_i^2 \lambda_j^2\rangle + 3 \langle\lambda_i^3 \lambda_j\rangle \end{displaymath}
\begin{displaymath} - \langle\lambda_i \lambda_j^2 \lambda_k\rangle) + \frac{1}{518400}\left(  - 3 \langle\lambda^5\rangle - 3 \langle\lambda_i \lambda_j^4\rangle  \right. \end{displaymath}\begin{equation}\left.- 4 \langle\lambda_i^3 \lambda_j^2\rangle + \langle\lambda_i \lambda_j^3 \lambda_k\rangle + \frac{1}{2} \langle\lambda_i^2 \lambda_j^2 \lambda_k\rangle \right).\end{equation}
\normalsize

As indicated in the text, we select all the triplets of roots that satisfy the constraint 
$\vec{\alpha}_i + \vec{\alpha}_j + \vec{\alpha}_k = 0$ and we evaluate on them the expression 
(A2), (A3) and (A4) appearing in Eq.(A1). The result is always zero, confirming the theorem of 
section  IV.

The combinations that satisfy the constraint  are, together with  generic permutations of 
$i, j, k$
\begin{enumerate}
\item $(c^a_i =1, c^a_j=1,c^a_k =-2)$ 
\item $(c^a_i =1, c^a_j=2,c^a_k =-3)$ 
\item $(c^a_i =1, c^a_j=3,c^a_k =-4)$ 
\item $(c^a_i =2, c^a_j=2,c^a_k =-4)$ 
\end{enumerate}
if $\lambda$ takes four different values,

\begin{enumerate}
\item $(c^a_i =1, c^a_j=1,c^a_k =-2)$ 
\item $(c^a_i =1, c^a_j=2,c^a_k =-3)$ 
\item $(c^a_i =1, c^a_j=3,c^a_k =-4)$ 
\item $(c^a_i =2, c^a_j=2,c^a_k =-4)$ 
\item $(c^a_i =1, c^a_j=4,c^a_k =-5)$ 
\item $(c^a_i =2, c^a_j=3,c^a_k =-5)$ 
\end{enumerate}
if $\lambda$ takes five different values and

\begin{enumerate}
\item $(c^a_i =1, c^a_j=1,c^a_k =-2)$ 
\item $(c^a_i =1, c^a_j=2,c^a_k =-3)$ 
\item $(c^a_i =1, c^a_j=3,c^a_k =-4)$ 
\item $(c^a_i =2, c^a_j=2,c^a_k =-4)$ 
\item $(c^a_i =1, c^a_j=4,c^a_k =-5)$ 
\item $(c^a_i =2, c^a_j=3,c^a_k =-5)$ 
\item $(c^a_i =1, c^a_j=5,c^a_k =-6)$ 
\item $(c^a_i =2, c^a_j=4,c^a_k =-6)$ 
\item $(c^a_i =3, c^a_j=3,c^a_k =-6)$ 
\end{enumerate}
if $\lambda$ takes six different values.


\begin{thebibliography}{99}
 \bibitem{'tHP} G.~'t~Hooft, \emph{Gauge theories for strong interactions}, in A.~Zichichi (ed.) 
``New phenomena in Subnuclear Physics'', Plenum Press, New York (1977).
 \bibitem{m} S.~Mandelstam, Phys. Rep. {\bf 23}, 245 (1976).
 \bibitem{'tH} G.~'t~Hooft, Nucl. Phys. B {\bf 79}, 276 (1974).
 \bibitem{Pol} A.~M.~Polyakov, JETP Lett.  {\bf 20}, 194 (1974).
 \bibitem{'tH2} G.~'t~Hooft, Nucl. Phys. B {\bf 190}, 455 (1981).
 \bibitem{dgt} T.~A.~DeGrand, D.~Toussaint, Phys. Rev. D {\bf 22}, 2478 (1980).
 \bibitem{sch} A.~S.~Kronfeld, M.~L.~Laursen, G.~Schierholz, U.~J.~Wiese, Phys. Lett. B {\bf 198}, 516 (1987).
 \bibitem{suz} T.~Suzuki, I.~Yotsuyanagi, Phys. Rev. D {\bf 42}, 4257 (1990).
 \bibitem{pol} M.~I.~Polikarpov, Nucl. Phys. B (Proc. Suppl.) {\bf 53}, 134 (1997), arXiv:hep-lat/9609020v1.
 \bibitem{CM} S.~R.~Coleman, J.~Mandula, Phys. Rev. {\bf 159}, 1251 (1967).
 \bibitem{abprog} L.~Del~Debbio, A.~Di~Giacomo, B.~Lucini, G.~Paffuti, arXiv:hep-lat/0203023.
 \bibitem{DLP} A.~Di~Giacomo, L.~Lepori and F.~Pucci, JHEP {\bf 0810}, 096 (2008), arXiv:0810.4226v1.
 \bibitem{arafune} J.~Arafune, P.~G.~O. Freund, C.~J.~Goebel, J. Math. Phys. {\bf 16}, 433 (1975).
 \bibitem{shnir} Ya.~Shnir, Magnetic Monopoles, Springer (2005).
 \bibitem{coleman} S.~Coleman, \emph{The Magnetic Monopole Fifty Years Later}, in A.~Zichichi (ed.) ``The unity of the fundamental interactions'', Plenum Press, New York (1983).
  \bibitem{gw} J.~Greensite, J.~Winchester, Phys. Rev. D {\bf 40}, 4167 (1989). 
 \bibitem{dmo} A.~Di~Giacomo, M.~Maggiore, S.~Olejnik, Nucl. Phys. B {\bf 347}, 441 (1990).
 \bibitem{sco}  T.~Sekido, K.~Ishiguro,Y.~Koma, Y.~Mori, T.~Suzuki, Phys. Rev. D {\bf 76}, 031501(R) (2007), arXiv:hep-lat/0703002v1.
 \bibitem{scol} T.~Suzuki, M.~Hasegawa, K.~Ishiguro, Y.~Koma, T.~Sekido, Phys. Rev. D {\bf 80}, 054504 (2009), arXiv:0907.0583v2.
 \bibitem{ddmo} L.~Del~Debbio, A.~Di~Giacomo, M.~Maggiore, S.~Olejnik, Phys. Lett. B {\bf267}, 254 (1991).
 \bibitem{dp} A.~Di~Giacomo, G.~Paffuti, Phys. Rev. D {\bf 56}, 6816 (1997), arXiv:hep-lat/9707003v1.
 \bibitem{ddp} L.~Del~Debbio, A.~Di~Giacomo, G.~Paffuti, Phys. Lett. B {\bf 349}, 513 (1995).
 \bibitem{ddpp} L.~Del~Debbio, A.~Di~Giacomo, G.~Paffuti, P.~Pieri, Phys. Lett. B {\bf 355}, 255 (1995), arXiv:hep-lat/9505014v1.
 \bibitem{dlmp} A.~Di~Giacomo, B.~Lucini, L.~Montesi, G.~Paffuti. Phys. Rev. D {\bf 61}, 034503 (2000), 	arXiv:hep-lat/9906024v2.
 \bibitem{ccos} P.~Cea, L.~Cosmai, Phys. Rev. D {\bf 62}, 094510 (2000), arXiv:hep-lat/0006007v1.
 \bibitem{3} J.~M.~Carmona, M.~D'Elia, A.~Di~Giacomo, B.~Lucini, G.~Paffuti, Phys. Rev. D {\bf 64}, 114507 (2001), arXiv:hep-lat/0103005v1.
 \end{thebibliography}
\end{document}